\documentclass[twocolumn,aps,floatfix,draft]{revtex4} 
\usepackage{amssymb}
\usepackage{amsfonts}
\newcommand{\trace}{\mathop{\rm Tr}\nolimits}

\newcommand{\bra}[1]{\langle#1|}
\newcommand{\ket}[1]{|#1\rangle}

\newcommand{\MAP}[1]{{\mathbf{#1}}}
\newcommand{\qed}{\hfill$\square$\par\vskip12pt} 

\newcommand{\cB}{{\cal B}}

\newcommand{\cH}{{\cal H}}

\newcommand{\cS}{{\cal S}}

\newcommand{\sP}{{\mathsf{P}}}
\newcommand{\sS}{{\mathsf{S}}}

\newcommand{\C}{{\mathbb{C}}}

\newcommand{\N}{{\mathbb{N}}}
\newcommand{\identity}{{\openone}} 

\newcommand{\be}{\begin{equation}}
\newcommand{\ee}{\end{equation}}
\newcommand{\bea}{\begin{eqnarray}}
\newcommand{\eea}{\end{eqnarray}}
\newcommand{\beas}{\begin{eqnarray*}}
\newcommand{\eeas}{\end{eqnarray*}}
\newtheorem{theorem}{Theorem} 
\begin{document}
\title{There, and Back Again:\\
Quantum Theory and Global Optimisation} 
\author{Koenraad M.R. Audenaert}
\affiliation{University of Wales, Bangor,
School of Informatics,
Bangor (Gwynedd) LL57 1UT, Wales}
\email{kauden@informatics.bangor.ac.uk}
\date{\today}
\begin{abstract}
We consider a problem in quantum theory that can be formulated as an optimisation problem and
present a global optimisation algorithm for solving it, the foundation of which relies in turn
on a theorem from quantum theory.
To wit, we consider the maximal output purity $\nu_q$ of a quantum channel as measured by Schatten $q$-norms, for integer $q$.
This quantity is of fundamental importance in the study of quantum channel capacities in quantum information theory.
To calculate $\nu_q$ one has to solve a non-convex optimisation problem that typically exhibits local optima.
We show that this particular problem can be approximated to arbitrary precision by an eigenvalue problem over a larger
matrix space, thereby circumventing the problem of local optima. The mathematical proof behind this algorithm relies
on the Quantum de Finetti theorem, which is a theorem used in the study of the foundations of quantum theory.

We expect that the approach presented here can be generalised and will turn out to be applicable to a larger class
of global optimisation problems.
We also present some preliminary numerical results, showing that, at least for small problem sizes, the present approach
is practically realisable.
\end{abstract}
\maketitle
%
\section{Introduction}
In Quantum Information Theory (QIT), noisy communication channels are modelled as completely positive trace-preserving
maps between operator algebras. One of the most fundamental questions in QIT is the determination of the capacity of
these quantum channels to transmit classical information, the so-called classical capacity of a quantum channel
\cite{holevo}.
What makes the
determination of this quantity so much more difficult than its purely classical counterpart is the existence of
entanglement in quantum physics. In fact, it has been shown that to obtain an optimal quantum channel decoder one
has to perform entangled measurements over the channel output states.
What is not known, however, is whether entanglement is also necessary to
obtain an optimal encoder. This is widely believed not to be the case, i.e.\ no benefit is expected
in having entanglement between the single-letter states.
To prove this, it is necessary to show that the single-letter
classical capacity (a.k.a.\ the Holevo capacity $\chi$) of a quantum channel is additive.

The Holevo capacity of a channel $\Phi$ is defined by
$$
\chi(\Phi) = \sup_{\pi,\rho} S(\sum \pi_i\Phi(\rho_i)) - \sum\pi_i S(\Phi(\rho_i)),
$$
where the sup runs over all probability distributions $\{\pi_i\}$ and collections of states $\{\rho_i\}$, and
$S$ is the von Neumann entropy $S(\rho) = -\trace[\rho\log\rho]$.
The additivity of this capacity would be the property $\chi(\Phi_1\otimes\Phi_2) = \chi(\Phi_1)+\chi(\Phi_2)$, i.e.
there would be no benefit in sending entangled states through the tensor product channel.

The expression for the Holevo capacity looks quite complicated, and it has resisted any attempt so far at
proving its additivity. For that reason,
a simpler channel property, called the minimal output entropy (MOE) \cite{amosov00}, has been introduced, in the hope that proving
additivity for the MOE would be simpler and yet shed some light on the additivity problem for the Holevo capacity.
The minimal output entropy $\nu_S$ of a channel $\Phi$ is defined as
$$
\nu_S(\Phi) = \min_\rho S(\Phi(\rho)),
$$
where the minimum runs over all density matrices (i.e. unit trace positive semidefinite matrices).
The entropy is minimal for those output states that are closest to being pure (rank 1).
It also makes sense to consider the related quantities
$$
\nu_q(\Phi) = \max_\rho ||\Phi(\rho)||_q,
$$
where purity is measured by the Schatten $q$-norm. These quantities are called the maximal output purities (MOP) and
they are directly related to the MOE:
$$
\nu_S(\Phi) = \lim_{q\downarrow 1} \frac{1-\nu_q^q(\Phi)}{q-1}.
$$
Furthermore, if the MOP is multiplicative, $\nu_q(\Phi_1\otimes\Phi_2) = \nu_q(\Phi_1) \nu_q(\Phi_2)$,
or if it is at least for values of $q$ close to 1, then the additivity of the MOE follows as a direct consequence.

Quite surprisingly, it has been shown recently that the additivity of the MOE is actually equivalent with the additivity
of the Holevo capacity \cite{Shor2}, in spite of the apparent greater simplicity of the expression for the MOE.
What is more, additivity of the MOE is also equivalent with the additivity of the Entanglement of Formation
\cite{Shor2,kasb}, another open problem in QIT whose resolution is eagerly awaited.
Additivity of the MOE and multiplicativity of the MOP have been proven in specific case \cite{AH,kingruskai1,king1,king2}.
Unfortunately, there is a counterexample to the multiplicativity of the MOP for values of $q>4.79$ \cite{werner02},
quenching the hope for a relatively simple proof of additivity of the MOE.
Nevertheless, MOP might still be multiplicative for smaller values of $q$.
Intuition has it that it might hold for $q$ between 1 and 2.
Since the Schatten $q$-norms are easier to work with for integer values of $q$, it is reasonable to first try and
prove multiplicativity of $\nu_2$ \cite{kingruskai3}.
If $\nu_2$ would indeed be multiplicative this would increase our belief in the above intuition.

In the following, we study the quantity $\nu_q$ for integer values of $q$, and the value 2 in particular.
The actual calculation of $\nu_q$ is an optimisation problem that is not very well-behaved, in the sense
that it exhibits local optima. Indeed, it involves the maximisation of a convex function over a convex set,
and the only reasonable statement that convex analysis can make is that the maxima will be obtained in extremal
points, i.e.\ pure states.
Our main technical result is that the optimisation problem defining $\nu_q$ can for integer $q$
be approximated by an eigenvalue problem
over a larger Hilbert space. This approximation yields an upper bound and the error of the approximation
goes to zero when the dimension of the enlarged Hilbert space goes to infinity.
\section{Main Results}
Consider a $d$-dimensional Hilbert space $\cH$, with its associated state space $\cS(\cH)$.
We will denote the set of bounded Hermitian operators over $\cH$ by $\cB^H(\cH)$.
Thus $\cS(\cH)$ is the subset of $\cB^H(\cH)$ containing all unit trace positive semidefinite operators.

In the following, we will consider the quantity $\nu_q^q(\Phi)=\max_{\rho\in\cS(\cH)} \trace[(\Phi(\rho))^q]$,
for integer $q$.
The maximand can be written as $\trace[\Phi(\rho)\Phi(\rho)\ldots\Phi(\rho)]$, with $q$ factors.
To every completely positive map $\Phi$ we can associate a positive semidefinite block matrix $\MAP{\Phi}$,
called the Choi matrix, such that $\Phi(\rho)_{ij} = \sum_{k,l}\MAP{\Phi}^{ij}_{kl}\rho_{kl}$.
Hence, the maximand can be brought in the general form $\trace[A \rho^{\otimes q}]$, where $A$ is a matrix
in $M_d(M_d(...M_d(\C)...))$ (with $q$ occurrences of $M_d$) depending on $\Phi$.
This matrix $A$ has composite row and column indices which we will denote by
$(i):=(i_1,i_2,\ldots,i_q)$ and $(j)$.
We will think of $A$ as being an operator over the tensor product $\cH^{\otimes q}$,
which is a Hilbert space consisting of $q$ copies of $\cH$.
Explicitly, we have
\be \label{eq:map}
A_{(i),(j)} = \trace[\MAP{\Phi}_{i_1,j_1} \ldots \MAP{\Phi}_{i_q,j_q}].
\ee
For $q=2$, $A$ is Hermitian, for higher $q$ this is generally not true.
\subsection{Symmetry}
Consider permutations of $n$ copies of $\cH$, $\pi\in S_n$, where $S_n$ is the symmetric group of order $n$.
If, for every permutation $\pi\in S_n$, a matrix $A$ over $\cH^{\otimes n}$ obeys
$A_{(i_1,\ldots,i_q),(j_1,\ldots,j_q)} = A_{(i_{\pi(1)},\ldots,i_{\pi(q)}),(j_{\pi(1)},\ldots,j_{\pi(q)})}$,
then the matrix $A$ is \textit{symmetric} (not to be confused with transposition symmetry).
To denote the action of a permutation $\pi\in S_n$ on a composite index,
we will use the abbreviation $\pi(i) :=(i_{\pi(1)},\ldots,i_{\pi(n)})$.
Let $P_\pi$ be the permutation matrix that permutes the indices according to $\pi$,
i.e.\ $(P_\pi x)_{(i)} = x_{\pi(i)}$. Thus $A$ is symmetric if and only if $\forall \pi\in S_n$,
$P_\pi^\dagger A P_\pi = A$.
The set of all symmetric Hermitian operators over $\cH^{\otimes n}$ will be denoted by $\cB^{HS}(\cH^{\otimes n})$.
This set is obviously a subspace of $\cB^{H}(\cH^{\otimes n})$.
The set of all symmetric states over $\cH^{\otimes n}$ will be denoted by $\cS^S(\cH^{\otimes n})$.
The state $\rho^{\otimes q}$ is a simple example of a symmetric state.
The linear map $\sP_n$ that projects all operators
in $\cB(\cH^{\otimes n})$ to $\cB^S(\cH^{\otimes n})$ is given by
$$
\sP_n(A) = \frac{1}{n!} \sum_{\pi\in S_n} P_\pi^\dagger A P_\pi.
$$
We call $\sP_n(A)$ the \textit{symmetric part} of $A$.

If, as in our case, the matrix $A$ is defined by the relation (\ref{eq:map})
where $\Phi$ is a completely positive map,
then $A$ is in general only Hermitian and symmetric for $q=2$.
However, $\sP_q(A)$ is Hermitian for all values of $q$ (and, of course, symmetric),
as follows easily from (\ref{eq:map}).

We will also need to consider the symmetry properties of vectors.
Vectors $\psi\in \cH^{\otimes n}$ are \textit{totally symmetric} if, for all $\pi\in S_n$,
$P_\pi \psi=\psi$, or $\psi_{(i)} = \psi_{\pi(i)}$.
The totally symmetric vectors form a subspace, which we will denote by $\sS(\cH^{\otimes n})$, and
which has dimension $S(d,n):=C^{n+d-1}_{d-1}$.
We denote by $\sS_n$ the projector on $\sS(\cH^{\otimes n})$, that is
$$
\sS_n = \frac{1}{n!}\sum_{\pi\in S_n} P_\pi.
$$
Finally, $P_n$ denotes the matrix whose columns span $\sS(\cH^{\otimes n})$ and are normalised
such that $P_n^\dagger P_n = \identity$.
\subsection{Main Theorem}
Let $\identity$ be the identity matrix over $\cH$.
Denote the maximal eigenvalue of a matrix $X$ and its corresponding eigenvector
by $\lambda_{\max}(X)$ and $\psi_{\max}(X)$.
\begin{theorem}
For any $q,n\in\N$, and for any $A\in\cB(\cH^{\otimes q})$
with Hermitian symmetric part ($\sP_q(A)\in\cB^H(\cH^{\otimes q})$),
there exists a non-increasing sequence $(\mu_n(A))_n$, with
$$
\mu_n(A):=\lambda_{\max}(\sP_{q+n}(A\otimes \identity^{\otimes n})),
$$
converging to
$$
\lim_{n\rightarrow \infty} \mu_n(A) = \max_\rho \trace[A \rho^{\otimes q}].
$$

Moreover, the optimal $\rho$ is given by
\beas
\rho_{\text{\rm opt}} &=& \lim_{n\rightarrow\infty}\trace_{q+n-1}\ket{\psi_n}\bra{\psi_n} \\
\psi_n &=& \psi_{\max}(\sP_{q+n}(A\otimes \identity^{\otimes n})).
\eeas
\end{theorem}
Note that the matrix $A$ need not be symmetric; one of the things the Theorem tells us is that
only its symmetric part is of relevance.

The main ingredient of the proof of this Theorem is the so-called Quantum de Finetti Theorem (QdF) \cite{stormer,hudson1},
which is a non-commutative analog of de Finetti's Theorem from probability theory.
The QdF theorem characterises the so-called \textit{exchangeable} sequences of states $(\rho^{(n)})_n$,
which are sequences of symmetric states over $\cH^{\otimes n}$
such that $\forall n,m\in\N: \rho^{(n)} = \trace_m[\rho^{(n+m)}]$.
Here $\trace_m$ denotes the partial trace over $m$ copies of $\cH$.
The QdF theorem states that a sequence is exchangeable if and only if there exists a positive measure $d\mu(\rho)$ over
$\cS(\cH)$ such that for all $n\in \N$:
$$
\rho^{(n)} = \int_{\cS(\cH)} \rho^{\otimes n} d\mu(\rho).
$$
The QdF theorem has been used in the study of the foundations of quantum mechanics \cite{hudson2,caves},
in mathematical physics \cite{fannes},
and recently also in QIT \cite{doherty}.
\subsection{Proof of Theorem 1.}
We are concerned with the maximisation of $\trace[A \rho^{\otimes q}]$ over all density matrices $\rho$.
We first turn this into the more general optimisation problem
$$
\max_{d\mu(\rho)\ge0} \{\trace[A \int \rho^{\otimes q} d\mu(\rho)]: \int d\mu(\rho) = 1\},
$$
which obviously achieves the same maximum value; the optimal measure $d\mu$ will be the Dirac measure peaking at
the optimal $\rho$.
We now claim that we can replace the latter maximisation by
a maximisation of $\trace[A \rho^{(q)}]$ over all $\rho^{(q)}$ that are $q$-th element in some exchangeable
sequence of states.
Indeed, any state of the form $\rho^{\otimes q}$ obviously forms part of an exchangeable sequence.
Conversely, by the QdF theorem,
the $q$-th element in any exchangeable sequence must be of the form $\int \rho^{\otimes q} d\mu(\rho)$
for some positive measure $d\mu(\rho)$.

From the definition of exchangeable sequence, we infer that $\rho^{(q)}$ must be the partial trace
of a symmetric state $\rho^{(\infty)}$ in $\cS^S(\cH^{\otimes\infty})$, where all but $q$ copies of
$\cH$ have been traced out.
Infinite direct products of Hilbert spaces, as well as of states, are mathematically well-defined.
The Hilbert space $\cH^{\otimes\infty}$ is the infinite direct
product of $\cH$, defined as an inductive limit \cite{stormer,takeda}.
The infinite direct product of a state $\rho$, $\rho^{\otimes\infty}$, is the unique state on
$\cH^{\otimes\infty}$ such that if
$A_i=\identity$ for all but a finite number $m$ of $A_i$,
$$
\trace[\rho^{\otimes\infty} \,\, (\otimes_i A_i)] = \prod_{j=1}^m \trace[\rho A_j].
$$

As a result, we can express the states $\int \rho^{\otimes q} d\mu(\rho)$ as a partial trace
of symmetric states in $\cS^S(\cH^{\otimes\infty})$:
\beas
\lefteqn{\max_\rho \trace[A \rho^{\otimes q}]} \\
&=& \lim_{n\rightarrow \infty} \max_{\rho} \{\trace[A \, \trace_{n}[\rho]]: \rho\in\cS^S(\cH^{\otimes(q+n)})\} \\
&=& \lim_{n\rightarrow \infty} \max_{\rho} \{\trace[(A\otimes\identity^{\otimes n}) \, \rho]:
\rho\in\cS^S(\cH^{\otimes(q+n)})\}.
\eeas
This maximisation over symmetric states can be replaced by a maximisation over all states,
provided they are projected first onto the symmetric subspace:
\beas
\lefteqn{\max_\rho \trace[A \rho^{\otimes q}]} \\
&=& \lim_{n\rightarrow \infty} \max_{\rho} \{\trace[(A\otimes\identity^{\otimes n}) \, \sP_{q+n}(\rho)]:
\rho\in\cS(\cH^{\otimes(q+n)})\}.
\eeas
The cyclicity of the trace of a matrix product implies that the projection can equally well be applied to
the factor $A\otimes\identity^{\otimes n}$, yielding
\bea
\lefteqn{\max_\rho \trace[A \rho^{\otimes q}]} \nonumber \\
&=& \lim_{n\rightarrow \infty} \max_{\rho} \{\trace[\sP_{q+n}(A\otimes\identity^{\otimes n}) \, \rho]:
\rho\in\cS(\cH^{\otimes(q+n)})\} \nonumber \\
&=& \lim_{n\rightarrow \infty} \lambda_{\max}(\sP_{q+n}(A\otimes\identity^{\otimes n})). \label{eq:1}
\eea

To show that $\lambda_{\max}(\sP_{q+n}(A\otimes\identity^{\otimes n}))$ is non-increasing
with $n$, we use the convexity of $\lambda_{\max}$ over $\cB^H$:
\beas
\lefteqn{\lambda_{\max}(\sP_{q+n+1}(A\otimes\identity^{\otimes (n+1)}))} \\
&=& \lambda_{\max}(\sP_{q+n+1}(\sP_{q+n}(A\otimes\identity^{\otimes n})\otimes\identity) \\
&\le& \frac{1}{(q+n+1)!}\sum_{\pi\in S_{q+n+1}}
\lambda_{\max}(P_\pi^\dagger(\sP_{q+n}(A\otimes\identity^{\otimes n})\otimes\identity)P_\pi) \\
&=& \lambda_{\max}(\sP_{q+n}(A\otimes\identity^{\otimes n})\otimes\identity) \\
&=& \lambda_{\max}(\sP_{q+n}(A\otimes\identity^{\otimes n})).
\eeas
This proves that the convergence of the right-hand side of eq.\ (\ref{eq:1}) is monotonically decreasing, so that
the right-hand side of (\ref{eq:1}) is an upper bound on the left-hand side of (\ref{eq:1})
for all finite values of $n$.

From the above calculations, we see that the optimal state $\rho$ in $\max_\rho \trace[A \rho^{\otimes q}]$
is approximated by
$$
\rho_{\text{opt}}^{\otimes q}
\approx \trace_n[\sP_{q+n}(\ket{\psi_n}\bra{\psi_n})]
= \trace_n\ket{\psi_n}\bra{\psi_n},
$$
with $\psi_n=\psi_{\max}(\sP_{q+n}(A\otimes\identity^{\otimes n}))$.
Therefore,
$$
\rho_{\text{opt}} = \lim_{n\rightarrow\infty} \trace_{q+n-1}\ket{\psi_n}\bra{\psi_n}.
$$
\qed
\subsection{Simplifications for the Calculation of $\nu_q(\Phi)$}
In this section we will show that in actual calculations of $\nu_q(\Phi)$ we do not have to deal with the (very large)
space $\cH^{\otimes(q+n)}$. Henceforth we will assume that $A$ corresponds to $\nu_q(\Phi)$, via
the relation (\ref{eq:map}). Furthermore, for $q>2$, we will assume that $A$ is already the symmetric part
of the originally obtained matrix. Thus, henceforth, $A$ is symmetric and Hermitian.

The first thing we note is that the symmetrisation $\sP_{q+n}$ in $\sP_{q+n}(A\otimes\identity^{\otimes n})$,
which in general involves a sum over all $(q+n)!$ permutations, can be replaced by a symmetrisation
over only $C^{q+n}_q$ permutations, provided $A$ is symmetric (or is made so, using $\sP_q$).
Indeed, both factors $A$ and $\identity^{\otimes n}$ already are symmetric, hence the only permutations necessary are
those that redistribute the $q$ indices of $A$ among the $q+n$ indices (considering the indices of $A$ to be
indistinguishable).

The most important issue here is to simplify the calculation of the maximal eigenvalue of
$\sP_{q+n}(A\otimes\identity^{\otimes n})$,
and we can actually do that by exploiting the full permutation symmetry of that matrix.
It is well-known that the eigenspaces of a fully symmetric matrix have to be themselves symmetry-invariant.
Indeed, if $A$ is symmetric then $\forall \pi\in S_n: P_\pi^\dagger A P_\pi = A$.
Inserting this in the eigenvalue equation $Ax=\lambda x$ gives $P_\pi^\dagger A P_\pi x=\lambda x$,
or, using unitarity of $P_\pi$, $A P_\pi x=\lambda P_\pi x$. In other words, $\forall \pi\in S_n$,
$P_\pi x$ is in the same eigenspace as $x$. It follows that every eigenspace is an $S_n$-invariant subspace
of $\cH^{\otimes n}$.

The irreducible $S_n$-invariant subspaces correspond to the symmetry classes of $S_n$. The best-known
are the totally symmetric and the totally antisymmetric class, but, of course, there are many others,
each one corresponding to a specific standard Young tableau of $S_n$.
The dimension of an $S_n$-invariant subspace depends on the dimension $d$ of the underlying
Hilbert space $\cH$. For the totally symmetric subspace, the dimension is $C^{d+n-1}_n$, and
for the totally antisymmetric subspace it is $C^d_n$, which is zero if $n>d$.
In fact, the dimension of any invariant subspace corresponding to a Young tableau of height larger than $d$
is zero, so we only need to consider Young tableaux with height at most $d$.
Still, the total number of invariant subspaces for a given $n$ grows exponentially with $n$.

If $U$ is a unitary matrix (depending on $d$ and $n$ only)
whose columns are basis vectors for the irreducible $S_n$-invariant
subspaces, then for a symmetric $A$, $U^\dagger AU$ is block-diagonal, every block corresponding to one of the
invariant subspaces.
Likewise, for general $A$, $\sP_n(A)$ is unitarily equivalent with the matrix consisting of the direct sum
of all the diagonal blocks of $U^\dagger AU$.
So, in order to calculate the maximal eigenvalue of a symmetric matrix, one needs only to calculate the
maximal eigenvalues of each of the mentioned diagonal blocks, and then take the maximum of those.

For every symmetry class, therefore, one can construct a matrix $U_k$ (for the $k$-th symmetry class)
whose columns are basis vectors of the corresponding invariant subspace, so that
$U_k^\dagger A U_k$ is the corresponding diagonal block.
Moreover, if $A$ is not symmetric, $U_k^\dagger A U_k$ is the corresponding diagonal block
for $\sP_n(A)$. Hence, in practice, the symmetrisation operation $\sP_n$ need not be performed at all.
This is good news, but the fact that the number of symmetry classes grows exponentially with $n$ is still
a nuisance.

Quite fortunately,
when we specialise to the case we are really interested in, namely the calculation of the maximal output purity
of a channel, we only need to consider the totally symmetric subspace.
Recall that $\nu_q^q(\Phi) = \max_\rho \trace[\Phi(\rho)^q] = \max_\rho \trace[A\rho^{\otimes q}]$.
The function $\trace[\Phi(\rho)^q]$ is convex, so its maximum over the state space will be achieved in a
pure state $\rho$. Hence $\trace[A\rho^{\otimes q}]$ will also be maximal for a pure $\rho$.
On the other hand, Theorem 1 states that
$\rho_{\text{opt}}=\lim_{n\rightarrow\infty}\trace_{q+n-1}\ket{\psi_n}\bra{\psi_n}$
with $\psi_n = \psi_{\max}(\sP_{q+n}(A\otimes \identity^{\otimes n}))$.
The only way in which $\rho_{\text{opt}}$ can tend to a pure state, say $\psi_{\text{opt}}$, is if $\psi_n$
tends to a tensor product $\psi_{\text{opt}}\otimes \psi'$ (some $\psi'$; this will of course have to be a
tensor power of $\psi_{\text{opt}}$). As a consequence, $\psi_n$ must tend to a totally symmetric state.
So, $\lambda_{\max}(\sS_{q+n}(A\otimes \identity^{\otimes n})\sS_{q+n})$ tends to the correct solution (where
$\sS_{q+n}$ is the projector on the totally symmetric subspace).
Furthermore, it does so in a non-increasing way as well.
Since $\sS_{q+n}$ is a projector, it is a contraction.
This, together with the fact that $A$ is Hermitian here, is used for the inequality in:
\beas
\lefteqn{\lambda_{\max}(\sS_{q+n+1}(A\otimes \identity^{\otimes n}\otimes\identity)\sS_{q+n+1})} \\
&=& \lambda_{\max}(\sS_{q+n+1}(\sS_{q+n}(A\otimes \identity^{\otimes n}) \sS_{q+n} \otimes\identity)\sS_{q+n+1})\\
&\le& \lambda_{\max}(\sS_{q+n}(A\otimes \identity^{\otimes n}) \sS_{q+n} \otimes\identity) \\
&=& \lambda_{\max}(\sS_{q+n}(A\otimes \identity^{\otimes n}) \sS_{q+n}),
\eeas
and that is what we needed to show.
We conclude that for the case of the MOP
there is no point in considering other subspaces than the totally symmetric one.

A final simplification is now possible. To reduce the dimension of the calculation we of course do not
calculate $\sS_{q+n}(A\otimes \identity^{\otimes n}) \sS_{q+n}$ directly, but rather
$P_{q+n}^\dagger (A\otimes \identity^{\otimes n}) P_{q+n}$, which is a matrix over a $S(d,q+n)$-dimensional
Hilbert space $\cH'$ (recall that $P_n$ is the matrix whose columns
span the totally symmetric subspace $\sS(\cH^{\otimes n})$, while $\sS_{q+n}$ is the projector on it).
There is no actual need to compute this over $\cH^{\otimes(q+n)}$ directly, because
this calculation is just a linear mapping from $\cB(\cH^{\otimes q})$ to $\cB(\cH')$.

We first introduce some additional notations.
Recall that $\cH$ is a $d$-dimensional Hilbert space and
vectors in $\cH^{\otimes n}$ are indexed by $(i):=(i_1,\ldots,i_n)$.
Every $i_j$ thus takes integer values from 1 to $d$.
If $x$ is a symmetric vector in $\cH^{\otimes n}$, then for any $\pi\in S_n$,
$x_{(i)}=x_{\pi(i)}$. This symmetry induces an equivalence relation on the set of indices, $(i)\sim\pi(i)$,
and we will use as equivalence class representative for an $n$-dimensional index $(i)$
the $d$-dimensional index $[k]:=[k_1,\ldots,k_d]$
where $k_j$ is the number of times the value $j$ occurs in the index $(i)$.
We denote this by $[k]=\#(i)$. Obviously, $\sum[k] := \sum_{j=1}^d k_j = n$.
One easily sees that $[k]$ can assume $S(d,n)$ possible values. The size of the class
corresponding to $[k]$ is the value of the multinomial coefficient $C^n_{[k]} := n!/k_1!\ldots k_d!$.
We adopt the common convention to take the value of the multinomial coefficient to be 0
whenever one or more of the $k_j$ is negative.

Using these notations, an explicit form for $P_n$ is:
$$
(P_n)_{(i),[k]} = \delta_{\#(i),[k]} c_{[k]}.
$$
The normalisation constant $c_{[k]}$ can be obtained from the requirement $P_n^\dagger P_n=\identity$:
$$
c^2_{[k]} = \#\{(j):\#(j)=[k]\} = C^n_{[k]}.
$$
Here $\#Z$ denotes the cardinality of the set $Z$.
Thus
$$
(P_n)_{(i),[k]} = \delta_{\#(i),[k]} (C^n_{[k]})^{-1/2}.
$$

We will denote the quantity we want to calculate by
$$
Q_n(A) := P_{q+n}^\dagger (A\otimes \identity^{\otimes n}) P_{q+n}.
$$
Inserting the above expression for $P_{q+n}$ gives, with $(i)$ and $(j)$ now being $(q+n)$-dimensional indices,
\beas
Q_n(A)_{[k],[l]} &=& \sum_{(i),(j)}
\delta_{\#(i),[k]} \,\, \delta_{\#(j),[l]}
(C^{q+n}_{[k]} C^{q+n}_{[l]})^{-1/2} \\
&& \times A_{(i_1,\ldots,i_q),(j_1,\ldots,j_q)}
\,\, \delta_{i_{q+1},j_{q+1}} \ldots \delta_{i_{q+n},j_{q+n}} \\
&=& (C^{q+n}_{[k]} C^{q+n}_{[l]})^{-1/2} \sum_{(i'),(j')} A_{(i'),(j')} \\
&& \times \sum_{(i'')} \delta_{\#(i'i''),[k]} \,\, \delta_{\#(j'i''),[l]},
\eeas
where $(i')$ and $(j')$ are $q$-dimensional, $(i'')$ is $n$-dimensional, and
$(i'i'')$ denotes the concatenation of $(i')$ and $(i'')$.
Since $\#(ab)=\#(a)+\#(b)$, the sum over $(i'')$ reduces to
$$
\sum_{(i'')} \delta_{\#(i'i''),[k]} \,\, \delta_{\#(j'i''),[l]}
= C^n_{[k]-\#(i')} \,\, \delta_{[k]-\#(i'),[l]-\#(j')}.
$$
Thus, with $[u]$ and $[v]$ being $d$-dimensional indices whose sum equals $q$,
\beas
\lefteqn{(C^{q+n}_{[k]} C^{q+n}_{[l]})^{1/2}\,\,Q_n(A)_{[k],[l]}} \\
&=& \sum_{(i'),(j')} A_{(i'),(j')} C^n_{[k]-\#(i')} \,\, \delta_{[k]-\#(i'),[l]-\#(j')} \\
&=& \sum_{[u],[v]} C^n_{[k]-[u]} \,\, \delta_{[k]-[u],[l]-[v]} \\
&& \times \sum_{(i'),(j')} \{ A_{(i'),(j')}: \#(i')=[u], \#(j')=[v]\} \\
&=& \sum_{[u],[v]} C^n_{[k]-[u]} \,\, \delta_{[k]-[u],[l]-[v]} \,\, (C^q_{[u]} C^q_{[v]})^{1/2} \\
&& \times \,\,\, (P_q^\dagger A P_q)_{[u],[v]}.
\eeas
This is final result.

Note that the presence of the Kronecker delta in this formula implies
that $Q_n(A)$ will have the same number of diagonals as $P_q^\dagger A P_q$, which
is $2S(d,q)-1$. The number of rows of $Q_n(A)$ is $S(d,q+n)$, as we have noted before.
If we treat $Q_n(A)$ as a sparse matrix, its storage requirements
are therefore just under $(2S(d,q)-1)S(d,q+n)$.
In function of $n$, this is of the order $O(n^{d-1})$.
\subsection{Numerical Study}
While our interest in the characterisation of $\nu_q(\Phi)$ by Theorem 1 is mainly theoretical --- the hope is
that it will ultimately lead to proving (or disproving) multiplicativity of $\nu_2$ --- it is also clear
that Theorem 1 can be immediately converted to an algorithm for the global maximisation of
$\trace[A \rho^{\otimes q}]$. Since this is a non-convex function of $\rho$ it is most noteworthy
that we have an algorithm here that is impervious to local maxima, unlike so many other optimisation algorithms.

We have performed a preliminary numerical study on the convergence behaviour of the sequence $\mu_n(A)$, for
the calculation of the maximal output 2-purity $\nu_2$ of qubit channels ($d=2$ and $q=2$).
The matrix $Q_n(A)$ is in this case a banded matrix with 2 diagonals on both sides of the main diagonal.
In all our experiments, the approximation error $\mu_n(A)-\mu_\infty(A)$ was of the order
$O(1/n)$. In fact, it turned out that the approximation error in function of $n$ could itself be approximated
to very good precision by a function of the form $1/(an+b)$, so much so that it proved
possible to obtain a much better value of $\mu_\infty(A)$ by extrapolating from a finite sequence of $\mu_n(A)$ values.
This extrapolation technique is similar in spirit to Aitken's method for accelerating the convergence of
the power method for eigenvalue calculations.
The obtained accuracy was frequently of the order of $10^{-10}$ (depending on the particular $A$),
while the total running time was less than 1 second on a standard PC. More details on these experiments will
be presented elsewhere.

The time complexity of the algorithm is $O(n^2)$, as it is dominated by the calculation
of the maximal eigenvalue $\lambda_{\max}$ of a banded matrix of fixed width.
We used the Matlab routine {\tt eigs(X,1,'LR')} (which calls the ARPACK library)
to obtain the maximal eigenvalue of a Hermitian matrix $X$.
\section{Conclusion}
We have presented a global optimisation algorithm for a particular problem occurring in quantum information theory.
The algorithm is based on representing the optimisation problem as an eigenvalue problem over a larger space.
The key to this representation is the Quantum de Finetti theorem, which is quite surprising given that its typical usage
is in the study of the foundations of quantum mechanics.
A lot of work remains to be done, both on the theoretical side (given our interest in exploiting Theorem 1 for settling
the multiplicativity issues of $\nu_q$) and on the algorithmic one.
It could very well be that Theorem 1 is just the simplest case of a larger class of optimisation problems
that in a similar way can be transformed to an eigenvalue problem.
We also believe this work will be of relevance to the study of higher-order tensors \cite{lieven}.

\end{document}